\documentclass[letterpaper, 10 pt, conference]{ieeeconf}  

\IEEEoverridecommandlockouts                              
\usepackage{amsmath} 
\usepackage{amssymb}  
\usepackage{graphicx,cite}
\usepackage{algorithm,algorithmic}

\usepackage[caption=false,font=footnotesize]{subfig}

\newcommand{\hs}{& \hspace{-2mm}}

\newcommand{\mc}[1]{\mathcal{#1}}
\newcommand{\bR}{\mathbb{R}}

\usepackage{theorem}
\theorembodyfont{\rmfamily}

\newtheorem{lem}{Lemma}
\newtheorem{defin}{Definition}
\newtheorem{theorem}{Theorem}
\newtheorem{prop}{Proposition}
\newtheorem{assum}{Assumption}

\newtheorem{exa}{Example}

\renewcommand{\QED}{\QEDopen}

\title{\LARGE \bf
Green Routing Game: Strategic Logistical Planning\\ using Mixed Fleets of ICEVs and EVs
}

\author{Hampei Sasahara, Gy\"{o}rgy D\'{a}n, Saurabh Amin, and Henrik Sandberg
\thanks{This work was supported in part by the C3.ai Digital Transformation Institute.}
\thanks{H. Sasahara is with the Department of Systems and Control Engineering, School of Engineering, Tokyo Institute of Technology, Tokyo 152-8552, Japan
        {\tt\small sasahara@sc.e.titech.ac.jp}}%
\thanks{G. D\'{a}n is with the Division of Network and Systems Engineering, School of Electrical Engineering and Computer Science, KTH Royal Institute of Technology, Stockholm SE-100 44, Sweden
		{\tt\small gyuri@kth.se}}%
\thanks{S. Amin is with the Laboratory for Information and Decision Systems,  Massachusetts Institute of Technology, Cambridge, MA 02139, USA
		{\tt\small amins@mit.edu}}%
\thanks{H. Sandberg is with the Division of Decision and Control Systems, School of Electrical Engineering and Computer Science, KTH Royal Institute of Technology, Stockholm SE-100 44, Sweden
        {\tt\small hsan@kth.se}}%
}

\begin{document}

\maketitle
\thispagestyle{empty}
\pagestyle{empty}

\begin{abstract}

This paper introduces a ``green'' routing game between multiple logistic operators (players), each owning a mixed fleet of internal combustion engine vehicle (ICEV) and electric vehicle (EV) trucks.
Each player faces the cost of delayed delivery  (due to charging requirements of EVs) and a pollution cost levied on the ICEVs.
This cost structure models:
1) limited battery capacity of EVs and their charging requirement;
2) shared nature of charging facilities;
3) pollution cost levied by regulatory agency on the use of ICEVs.
We characterize Nash equilibria of this game and derive a condition for its uniqueness.
We also use the gradient projection method to compute this equilibrium in a distributed manner. 
Our equilibrium analysis is useful to analyze the trade-off faced by players in incurring higher delay due to congestion
at charging locations when the share of EVs increases versus a higher pollution cost when the share of ICEVs increases.
A numerical example suggests that to increase marginal pollution cost can dramatically reduce inefficiency of equilibria.
\end{abstract}

\section{INTRODUCTION}

The transportation sector is the largest contributor to greenhouse gas emissions worldwide.
In 2019, it accounted for 29\% of $\mathrm{CO_2}$ emissions in the EU~\cite{European2021National}.
A noticeable drop was observed after 2020 due to the impact of COVID-19; however emissions are expected to return to (or even exceed) the pre-pandemic levels as economic activity recovers steadily.
Both governments and private sector have proposed ambitious plans for decarbonizing the transportation sector.
For example, the EU proposes to cut $\mathrm{CO_2}$ emissions by at least 55\% by 2030 and to become climate neutral by 2050~\cite{European2020Stepping}.
Most of these plans propose aggressive adoption of electric vehicles (EVs).
Technologically, EVs have emerged as a serious alternative to conventional vehicles, namely, internal combustion engine vehicles (ICEVs)~\cite{Yong2015Review}.
However, the current state of EV technology still requires a significant fixed and operational costs as well as access to reliable and fast charging infrastructure.
Thus, to support low-carbon freight logistics, one needs to analyze the incentives of \emph{strategic} fleet owners for maintaining a sizable proportion of EVs in their operations.


Previous literature has focused on efficient delivery operations using both ICEVs and EVs (or mixed fleets) from the perspective of a centralized operator or a single logistics operator~\cite{Bektacs2011Pollution,Schneider2014Electric,Goeke2015Routing,Kocc2016Green,Shen2019Optimization}.
There papers adopt variants of the vehicle routing problem after incorporating the operational costs, environmental externalities, and service and battery charging time requirements~\cite{Asghari2021Green}.

This paper proposes a \emph{green routing game,} to analyze the strategic routing behavior of multiple logistics operators who own a mixed fleet of ICEVs and EVs and operate over a parallel network with shared charging facilities.
The standard formulations in traffic routing games incorporate the congestible nature of transportation facilities~\cite{Roughgarden2007Game,Krichene2014Stackelberg,Lazar2021Routing,Wu2021Value}.
This model has also been shown to be suitable in analyzing multiuser communication~\cite{Orda1993Competitive} and demand-side management in smart grids~\cite{Chen2014Autonomous,Jacquot2018Analysis}.
Our model is adapted to the situation when the cost of delay is incurred by the EVs at the charging stations;
this is in contrast to classical models in which the congestion cost models externalities imposed by other vehicles sharing the same route(s).
Further, we consider atomic splittable flows~\cite{Jacquot2018Routing} to model the fleet composition and routing decisions of individual logistic operators.

Analysis of equilibria is a basis for measuring inefficiency of selfish routing and designing effective economic mechanisms.
We first show that the game can admit multiple equilibria in contrast to standard routing games, the existence of which complicates the analysis.
To specify tractable situations, we investigate a condition under which equilibria are essentially unique.
It is shown that the uniqueness holds when the delayed delivery penalties are identical with respect to players.
To compute the equilibrium, we consider the distributed algorithm proposed in~\cite{Jacquot2018Analysis}.
However, the convergence analysis cannot be straightforwardly extended to our game because each iteration of the algorithm is not necessarily a contraction, and consequently, the standard fixed point theorem cannot apply.
We prove the convergence based on its variant adapted to averaged maps~\cite{Byrne2003Unified}.
Based on those results, we present a numerical example to discuss relationship between the pollution cost, the trade-off that players face, and inefficiency of equilibria.
Importantly, the example suggests that high marginal pollution cost can dramatically reduce the inefficiency.


The paper is organized as follows.
Sec.~\ref{sec:model} introduces the green routing game as a model of the pollution-aware strategic routing with congestion at public charging stations.
In Sec.~\ref{sec:ana}, an example with multiple equilibria is exhibited.
The example suggests a condition under which the equilibrium is unique, and we formally prove the uniqueness.
Sec.~\ref{sec:alg} provides a distributed algorithm based on the gradient projection method.
In Sec.~\ref{sec:num}, numerical examples are presented, and finally, Sec.~\ref{sec:conc} concludes and summarizes the paper.

\subsection*{Notation}

Denote by $\mathbb{N},\bR,\bR_+,$ and $\bR^N$ the sets of natural numbers, real numbers, nonnegative real numbers, and $N$-dimensional Euclidean space, respectively.
The identity mapping is denoted by $I$.
The transpose and the Euclidean norm of a column vector $x\in\bR^N$ are denoted by $x^{\sf T}$ and $\|x\|$, respectively.
The spectrum of a matrix $M$ is denoted by $\sigma(M)$.

\section{GREEN ROUTING GAME}
\label{sec:model}

\begin{figure}[t]
  \centering
  \includegraphics[width=0.98\linewidth]{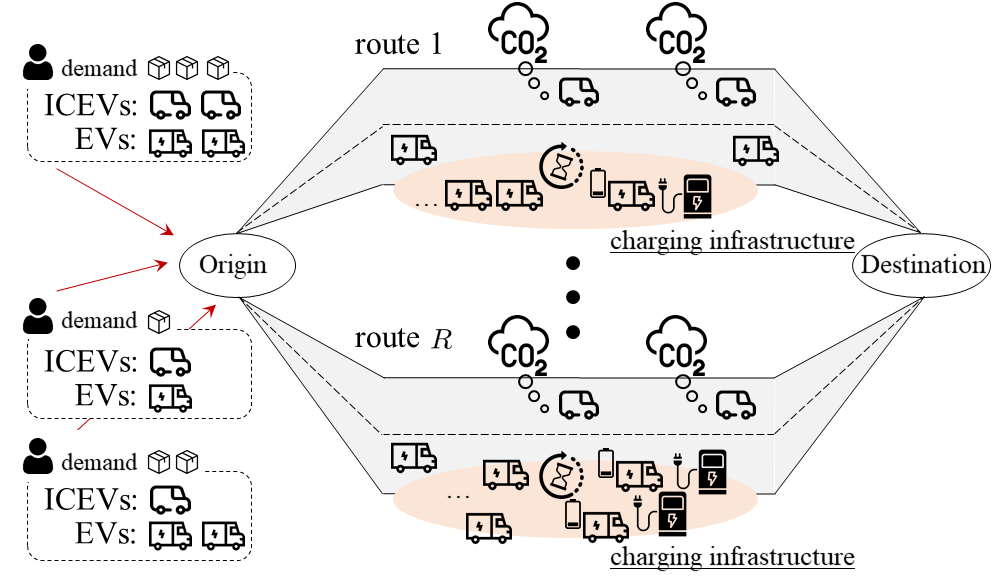}
  \caption{
  Situation of the green routing game.
  }
  \label{fig:des}
\end{figure}

We model strategic routing decisions by multiple logistic operators (players).
Each player's fleet consists of a mix of EVs and traditional ICEVs.
Players are required to deliver a predetermined amount of commodities (i.e. demand) from an origin node to a common destination node.
The players share a common EV charging infrastructure (specialized for heavy duty vehicles~\cite{Al2021Charging}) that is deployed over a parallel route network; see Fig.~\ref{fig:des}.
Every player decides the share of the demand to be delivered by the ICEVs and EVs as well as the route used by each vehicle in her fleet.
The player's cost is comprised of operational and environmental costs.
For sake of simplicity, we model these costs as the cost of delayed delivery at the destination node and pollution cost, respectively.
We also suppose that the congestion externality faced by the vehicle fleets due to presence of the regular traffic is negligible relative to the congestion faced by EVs at the charging stations.
In particular, the EVs are supposed to require access to charging stations to complete their trip due to the limitations of current battery technology.
Furthermore, the ``public'' nature of charging facilities implies that the delay incurred in waiting and charging at these locations must be accounted by the players in their routing decisions.
Thus, in our model, congestion only impacts the cost of delayed delivery of EVs.
Again, for simplicity, we consider that the ICEVs face no delays on any route.
However, players incur a pollution cost in the form of fuel and environmental tax for each ICEV that is routed through the network.
In contrast, the electricity supply at the charging stations is ``net-zero'' and hence the players do not incur pollution costs in routing EVs through the network.
The simplifying assumptions in our model are primarily governed by our intent to focus on the trade-off that the players face by incurring low cost of delayed delivery (resp. high pollution cost) when they choose a larger proportion of ICEVs instead of relaying on EVs that face congestion at charging stations.

Let $\mc{I}:=\{1,\ldots,N\}$ denote the set of players.
The amount of freight demand that needs to be delivered by player $i \in \mc{I}$ is given by $D^i\geq 0$.
Players share a network of parallel routes denoted by $\mc{R}:=\{1,\ldots,R\}$.
For simplicity, we assume that all ICEVs face an identical pollution cost regardless of the route assigned to them.
Moreover, without loss of generality, we consider that all EVs necessarily need to recharge en-route; otherwise delivering all demand by EVs using the route becomes the trivial optimal routing for the operator, who has no interaction among the other players.
We also consider that charging stations are placed at each route since roads without charging infrastructure cannot be exploited by EVs.
For player $i \in \mc{I}$, we denote the total demand delivered by ICEVs by $x^i_0\in\bR_+$ (note that we do not need to track the ICEVs in each route because of route-independent pollution cost in our model).
Similarly, we denote the demand delivered by EVs of player $i \in \mc{I}$ through route $r\in\mc{R}$ by $x^i_r\in\bR_+$.
Then the action of player $i$ is represented as
\[
 x^i:=(x^i_0,x^i_1,\ldots,x^i_R)\in\mc{X}^i,
\]
where the action set is given by
\[
 \textstyle{
  \mc{X}^i:=\{x^i\in\bR^{R+1}_{+}: x^i_0+\sum_{r\in\mc{R}}x^i_r=D^i\}.
 }
\]
The flow profile on the set of players is denoted by
$x:=(x^1,\ldots,x^N)$
and its feasible set is denoted by $\mc{X}:=\mc{X}^1\times\cdots\times\mc{X}^N$.
Denote the aggregated flow of EVs over each route by
\[
 \textstyle{X_r:=\sum_{i \in \mc{I}} x^i_r}.
\]

\begin{figure}[t]
  \centering
  \includegraphics[width=0.98\linewidth]{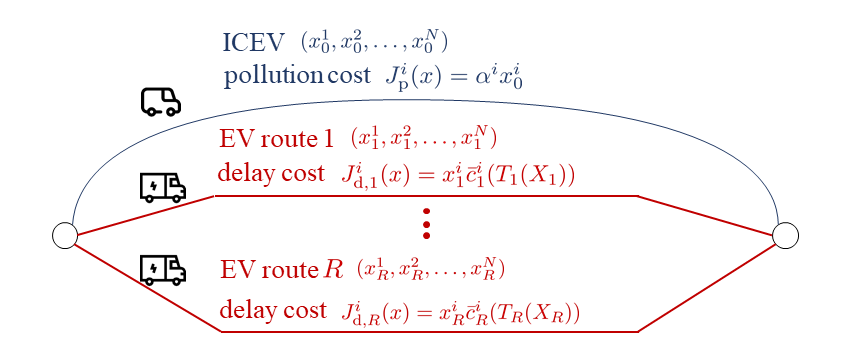}
  \caption{
  Mathematical ingredients of the green routing game.
  }
  \label{fig:model}
\end{figure}

Due to the delay incurred by EVs while waiting at the charging stations, the total expected delivery duration for each route depends on the aggregate flow that is routed through it.
We denote the delivery duration for route $r\in\mc{R}$ by $T_r(X_r)$ with $T_r:[0,\infty) \to [0,\infty)$.
On the other hand, the pollution cost linearly depends on the demand met by the ICEVs (although marginal cost of pollution can vary across players).
Fig.~\ref{fig:model} illustrates this setting.

The cost function of player $i$, denoted by $J^i:\mc{X}\to\bR_+$, is given by
\begin{equation}\label{eq:obj_fun}
 \textstyle{
  J^i(x)=J^i_{\rm p}(x)+\sum_{r\in\mc{R}}J^i_{{\rm d},r}(x)
 }
\end{equation}
where
\[
 J^i_{\rm p}(x):=\alpha^i x^i_0,\quad
 \textstyle{
  J^i_{{\rm d},r}(x):=x^i_r \bar{c}^i(T_r(X_r))
 }
\]
with $\alpha^i\geq 0$ and $\bar{c}^i:\bR_+\to\bR_+$ that satisfies
\begin{equation}\label{eq:no_pen_early_del}
 \bar{c}^i(t)=0,\quad \forall t\in[0,\tau^i]
\end{equation}
with some $\tau^i\geq0$.
In~\eqref{eq:obj_fun}, $J^i_{\rm p}:\mc{X}\to\bR_+$ and $J^i_{{\rm d},r}:\mc{X}\to\bR_+$ represent the pollution cost and the delayed delivery cost associated with the route $r\in\mc{R}$, respectively.
Here $\bar{c}^i$ represents the delayed delivery cost per unit time,
where the condition~\eqref{eq:no_pen_early_del} signifies that no cost is incurred for an early or on-time delivery.
In addition, we make the following technical assumptions:
For any $i \in \mc{I}$ and $r\in\mc{R}$,
\begin{itemize}
\item $\bar{c}^i(t)$ is strictly increasing and convex for $t\in[\tau^i,\infty)$.
\item $T_r(X)$ is strictly increasing and convex for $X\in[0,\infty)$.
\item $\bar{c}^i(t)$ and $T_r(X)$ are twice continuously differentiable for $t\in[0,\infty)$ and $X\in[0,\infty)$.
\end{itemize}
While $\bar{c}^i$ is determined from the operational requirement, $T_r$ is composed of driving and charging duration.
The driving duration is determined by the traffic congestion level, which can be estimated from data of GPS~\cite{Work2010Traffic} or road sensors~\cite{Polson2017Deep}.
The charging duration is determined by the capacity of the charging facility~\cite{Tomaszewska2019Lithium}.

\begin{figure}[t]
  \centering
  \includegraphics[width=0.98\linewidth]{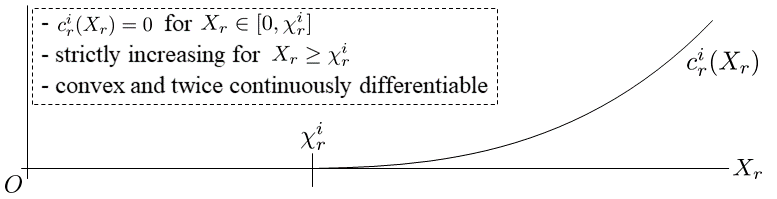}
  \caption{
  The form of the delayed delivery cost function $c^i_r$.
  }
  \label{fig:cir}
\end{figure}

In fact, we can simply express the delayed delivery cost as a function of the aggregated flow of each route by $c^i_r:X_r\mapsto \bar{c}^i(T_r(X_r))$.
Thus, we can write
\[
 \textstyle{
  J^i(x)=\alpha^i x^i_0 + \sum_{r\in\mc{R}} x^i_r c^i_r(X_r).
 }
\]
Similarly, the aforementioned assumptions on $\bar{c}^i_r$ and $T_r$ can be re-written as follows:
\begin{assum}\label{assum:tech}
For any $i \in \mc{I}$ and $r\in\mc{R}$,
\begin{itemize}
\item $c^i_r(X)=0$ for any $X\in[0,\chi^i_r]$ with some $\chi^i_r\geq0$.
\item $c^i_r(X)$ is strictly increasing for $X\in[\chi^i_r,\infty)$.
\item $c^i_r(X)$ is convex and twice continuously differentiable for $X\in[0,\infty)$.
\end{itemize}
\end{assum}
The form of $c^i_r$ is illustrated in Fig.~\ref{fig:cir}.
We refer to the interval $[0, \chi^i_r]$ as the \emph{cost-free interval} of the route $r\in\mc{R}$ for player $i\in\mc{N}$ and the route where the aggregated flow $X_r$ is within the cost-free interval as a \emph{cost-free route} given a flow profile.
Its existence immediately suggests that players prioritize cost-free routes, and moreover, induces multiple equilibria as observed in Sec.~\ref{sec:ana}.

The model leads to an $N$-player atomic routing game $\mc{G}:=(\mc{I},\{\mc{X}^i\}_{i \in \mc{I}},\{J^i\}_{i \in \mc{I}})$.
We adopt Nash equilibrium (NE) as the solution concept:
\begin{defin}[Nash Equilibrium]
A flow profile $x^{\ast}=(x^{1{\ast}},\ldots,x^{N{\ast}})\in\mc{X}$ is said to be a NE of the game $\mc{G}$ if
\[
 J^i(x^{\ast})\leq J^i(x^i,x^{-i{\ast}}),\quad \forall x^i\in\mc{X}^i
\]
for any $i \in \mc{I}$, where $J^i(x)=J^i(x^i,x^{-i})$ denotes the $n$-th player's cost function and $x^{-i}$ denotes the flow profile of all players but $i \in \mc{I}$.
\end{defin}

We say that the NE of $\mc{G}$ is essentially unique if the costs corresponding to all NE are equal.

Note that the existence of NE under Assumption~\ref{assum:tech} is straightforward from the well-known result: there exists a NE when every cost function is convex and every feasible region is compact and convex~\cite[Theorem 4.7.2]{Laraki2019Mathematical}.

\emph{Remark:} We assume the total cost $J^i(x)$ to be continuous with respect to the amount of the delivered commodities $x^i_r$.
In general, $J^i(x)$ can be discontinuous in $x$ because the total amount of commodity depends on the batches transported by individual vehicles.
Our model is obtained through smoothing on an implicit premise that the commodity amount delivered by a single vehicle is relatively small and the batch nature of transport can be ignored.

\section{EQUILIBRIUM ANALYSIS}
\label{sec:ana}

\subsection{Existence of Multiple NE}
\label{subsec:ex}

Analysis of NE is a basis for quantifying inefficiency of selfish routing and designing effective economic mechanisms.
When the NE is unique, the analysis becomes considerably simple.
For example, price of anarchy (PoA) and price of stability, which are different measures of inefficiency, are identical in a game with a unique NE~\cite[Chap.~17]{Roughgarden2007Game}.
However, the green routing game admits multiple NE in contrast to standard atomic splittable routing games over parallel networks~\cite{Orda1993Competitive,Jacquot2018Analysis}.
This subsection exhibits an example with multiple NE suggesting a condition for uniqueness of NE, which specifies tractable situations.



\begin{figure}[t]
  \centering
  \includegraphics[width=0.98\linewidth]{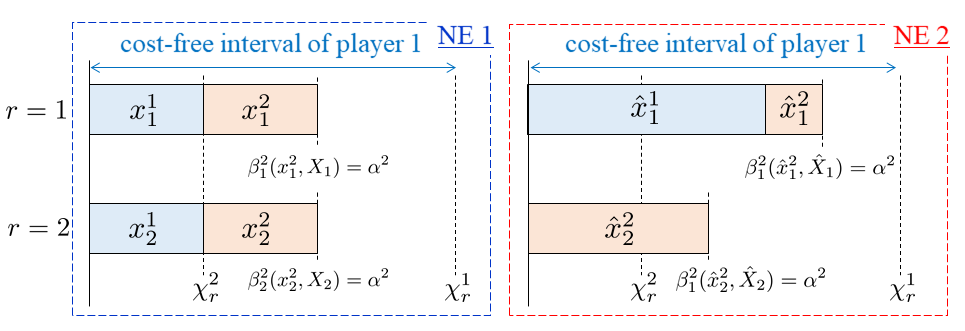}
  \caption{
  Example of multiple NE in a game with two players and two routes, where $x^1_0$ and $x^2_0$ are omitted.
  }
  \label{fig:multiNE_ex}
\end{figure}

\begin{exa}
Consider a 2-player game with two roads where $(D^1,D^2)=(2,10),$ $(\chi^1_r,\chi^2_r)=(1,10)$ for $r\in\{1,2\},$ and $(\alpha^1,\alpha^2)=(3,3)$.
The cost function over the cost-free interval is given by a quadratic function $c^i_r(X_r)=(X_r-\chi^i_r)^2$ for $i\in\{1,2\}$ and $r\in\{1,2\}$.
It can be shown that both
\[
\begin{array}{l}
 (x^1,x^2)=\left([0\ 1\ 1]^{\sf T}, [8\ 1\ 1]^{\sf T}\right),\\
 (\hat{x}^1,\hat{x}^2)=\left([0\ 2\ 0]^{\sf T}, [10-\frac{2\sqrt{10}}{3}\ \frac{-2+\sqrt{10}}{3}\ \frac{2+\sqrt{10}}{3}]^{\sf T}\right)
\end{array}
\]
are NE.
Moreover, the NE are not essentially unique.
In fact, the costs of player 2 corresponding to the NE are given by
\[
 \textstyle{
  J^2(x^1,x^2)=28,\quad J^2(\hat{x}^1,\hat{x}^2)\simeq 25.3.
 }
\]
The NE are illustrated in Fig.~\ref{fig:multiNE_ex}, where $\beta^i_r(x^i_r,X_r):=c^i_r(X_r)+x^i_rc^{i \prime}_r(X_r)$ denotes the marginal cost of the route $r$ for player $i$.
\end{exa}

The second NE is obtained as follows:
Since the aggregated flows are within the cost-free intervals for player 1 at the first NE, the equilibrium flow profile $(x^1,x^2)$ incurs no cost for player 1.
Also, since the aggregated flow $X_1$ is strictly less than the threshold $\chi^1_1$, moving a fraction of $x^1_2$ to the first route does not change the cost for player 1.
Thus, $\hat{x}^1$ is a best response to any routing of player 2 as long as $X_r\leq \chi^1_r$.
On the other hand, from the perspective of player 2, her reasonable action is changed by the lateral movement of player 1.
The reasonable usage of EVs at each route $\hat{x}^2_r$ is characterized by $\beta^2_r(\hat{x}^2_r,\hat{X}_r)=\alpha^2$, which means that the marginal cost of each route $\beta^2_r(\hat{x}^2_r,\hat{X}_r)$ is balanced with the marginal cost of usage of ICEVs given as $\alpha^2$.
This results in the second NE.
While the cost of player 1 is zero at both NE, the cost of player 2 varies, and hence those NE are not essentially unique.

This example suggests that the huge gap between the delayed delivery costs of the players causes the multiple NE.
It is expected that, conversely the NE can be unique with cost functions identical with respect to players.
The next subsection formally proves the uniqueness to specify tractable instances.

\subsection{Condition for Uniqueness of NE}



The following assumption is made for the subsequent analysis.
\begin{assum}\label{assum:identical_intervals}
The delayed delivery cost functions are identical with respect to the players, i.e.,
$c^1_r = c^2_r = \cdots = c^N_r$ for every $r\in\mc{R}$.
\end{assum}

In the following, we omit the superscript $i\in\mc{I}$ from $c^i_r$.
Practically, Assumption~\ref{assum:identical_intervals} holds when the operators provide the same service.
A possible situation is that a customer at the destination orders goods from multiple logistic companies imposing the same delay penalty.

In fact, Assumption~\ref{assum:identical_intervals} guarantees uniqueness of NE.
The following theorem holds.
\begin{theorem}\label{thm:uniqueness}
Let Assumptions~\ref{assum:tech} and~\ref{assum:identical_intervals} hold.
Then the NE of the game $\mc{G}$ are essentially unique.
\end{theorem}

To prove Theorem~\ref{thm:uniqueness}, we first introduce three structural results about the NE.
Note that, because the constraints associated with $\mc{X}^i$ satisfy linear independence constraint qualification, a flow profile $x\in\mc{X}$ is a NE if and only if it satisfies the Karush-Kuhn-Tucker (KKT) conditions~\cite[Chap.~5]{BoyVan}:
\begin{equation}\label{eq:KKT}
\begin{array}{l}
 \alpha^i=\lambda^i+\mu^i_0,\\
 \beta^i_r(x^i_r,X_r)=\lambda^i+\mu^i_r,\quad \forall r\in\mc{R},\\
 x^i_0+\sum_{r\in\mc{R}}x^i_r=D^i,\\
 x^i_r\geq 0,\quad \mu^i_rx^i_r=0,\quad \mu^i_r\geq 0, \quad \forall r\in \{0\}\cup \mc{R}
\end{array}
\end{equation}
with Lagrange multipliers $(\lambda^i,\mu^i_0,\mu^i_1,\ldots,\mu^i_r)\in\mathbb{R}^{R+2}$ for every $i \in \mc{I}$.

First, we claim that if there are cost-free routes at some NE, then all routes need to be cost-free at any NE.
\begin{lem}\label{lem:delay}
Let Assumptions~\ref{assum:tech} and~\ref{assum:identical_intervals} hold.
Assume that there exists a NE such that $X_r\leq\chi_r$ for some $r\in\mc{R}$.
Then every NE satisfies $X_r\leq\chi_r$ for any $r\in\mc{R}$.
\end{lem}
\begin{proof}
Note that Assumption~\ref{assum:identical_intervals} implies that $\chi^1_r=\chi^2_r=\cdots=\chi^N_r$ for any $r\in\mc{R}$.
We remove the superscript $i\in\mc{I}$ from $\chi^i_r$.
Let $x\in\mc{N}$ be the NE such that $X_r\leq\chi_r$ for some $r\in\mc{R}$.
We first prove that this NE satisfies $X_r\leq\chi_r$ for any $r\in\mc{R}$ by contradiction.
Assume that there exists $s\in\mc{R}$ such that $X_s > \chi_s$.
Then there exists $i\in\mc{I}$ such that $x^i_s>0$.
From the complementarity condition, $\mu^i_s=0$.
Thus $\beta^i_s(x^i_s,X_s)=\lambda^i$.
On the other hand, since $X_r\leq \chi_r$, $\beta^i_s(x^i_r,X_r)=\lambda^i+\mu^i_r=0$.
Therefore $\beta^i_s(x^i_s,X_s)=-\mu^i_r\leq 0$.
However, since $x^i_s>0$ and $X_s >\chi_s$, we have $\beta^i_s(x^i_s,X_s)>0$, which is a contradiction.

Finally, from the claim above, we have $X_r<\chi_r$ for every $r\in\mc{R}$.
Thus, for every feasible flow profile $\hat{x}\in\mc{X}$, which satisfies $\sum_{i\in\mc{I}}(\hat{x}^i_0+\sum_{r\in\mc{R}}\hat{x}^i_r)=\sum_{i\in\mc{I}}D^i$, there exists $r\in\mc{R}$ such that $\hat{X}_r<\chi_r$.
Therefore every NE satisfies the condition.
\end{proof}

Lemma~\ref{lem:delay} implies that if there exists a NE that causes free-cost intervals under Assumption~\ref{assum:identical_intervals} then the NE are essentially unique.
Thus, in what follows we can focus on NE at which there are no free-cost intervals.
In the remainder of this subsection, we denote two NE with the corresponding Lagrange multipliers by $(x,\lambda,\mu)$ and $(\hat{x},\hat{\lambda},\hat{\mu})$.
Also, the corresponding aggregated flows denoted by $X_r$ and $\hat{X}_r$ satisfy
\begin{equation}\label{eq:X_r}
 X_r>\chi_r,\quad \hat{X}_r > \chi_r,\quad \forall r\in\mc{R}.
\end{equation}
Under this condition, we basically follow the proof of~\cite[Theorem 2]{Jacquot2018Analysis}.
The following lemma holds.
\begin{lem}\label{lem:ineq}
Let Assumptions~\ref{assum:tech} and~\ref{assum:identical_intervals} hold.
If~\eqref{eq:X_r} holds, then
the following holds:
\begin{itemize}
\item If $\hat{\lambda}^i\leq \lambda^i$ and $\hat{X}_r\geq X_r$, then $\hat{x}^i_r\leq x^i_r$,
\item if $\hat{\lambda}^i\geq \lambda^i$ and $\hat{X}_r\leq X_r$, then $\hat{x}^i_r\geq x^i_r$,
\item if $\hat{\lambda}^i< \lambda^i$ and $\hat{X}_r\geq X_r$, then $\hat{x}^i_r< x^i_r$ or $\hat{x}^i_r=x^i_r=0$,
\item if $\hat{\lambda}^i> \lambda^i$ and $\hat{X}_r\leq X_r$, then $\hat{x}^i_r> x^i_r$ or $\hat{x}^i_r=x^i_r=0$,
\end{itemize}
and
\begin{itemize}
\item if $\hat{\lambda}^i<\lambda^i$ then $\hat{x}^i_0=0$,
\item if $\hat{\lambda}^i>\lambda^i$ then $x^i_0=0$.
\end{itemize}
\end{lem}
\begin{proof}
We show the first claim.
Assume $\hat{\lambda}^i\leq \lambda^i$ and $\hat{X}_r\geq X_r$.
If $\hat{\mu}^i_r>0$ then $\hat{x}^i_r=0$ from the complementarity condition, and hence $\hat{x}^i_r\leq x^i_r$.
If $\hat{\mu}^i_r=0$ then $\beta^i_r(\hat{x}^i_r,\hat{X}_r)=\hat{\lambda}^i\leq \lambda^i\leq\lambda^i+\mu^i_0=\beta^i_r(x^i_r,X_r)$.
Since $\beta^i_r$ is monotonically non-decreasing in $X_r$, $\beta^i_r(\hat{x}^i_r,\hat{X}_r)\geq \beta^i_r(x^i_r,X_r) \geq \beta^i_r(x^i_r,\hat{X}_r)$.
Since $\hat{X}_r>\chi^i_r$ from Lemma~\ref{lem:delay}, $c^{i \prime}_r(\hat{X}_r)>0$.
Hence $\beta^i_r(x^i_r,\hat{X}_r)$ is strictly increasing in $x^i_r$.
Thus $\hat{x}^i_r\leq x^i_r$.
Next, assume $\hat{\lambda}^i< \lambda^i$ and $\hat{X}_r\geq X_r$.
If $\hat{\mu}^i_r>0$ then $\hat{x}^i_r=0,$ which implies $\hat{x}^i_r<x^i_r$ or $\hat{x}^i_r=x^i_r=0$.
If $\hat{\mu}^i_r=0$ then $\beta^i_r(\hat{x}^i_r,\hat{X}_r)<\beta^i_r(x^i_r,X_r)$.
As in the deduction above, we have $\hat{x}^i_r<x^i_r$.
Finally, the second and fourth items can be proven in the same manner.

In terms of the second claim, if $\hat{\lambda}^i<\lambda^i$ then $\hat{\mu}^i_0>\mu^i_0$ from~\eqref{eq:KKT}.
Thus the complementarity condition leads to $\hat{x}^i_0=0$.
The second item can be proven in the same manner.
\end{proof}

Based on the preparation, we can show that the aggregated flow profiles of two NE are identical.
The following lemma holds.
\begin{lem}\label{lem:Xr}
Let Assumptions~\ref{assum:tech} and~\ref{assum:identical_intervals} hold.
If~\eqref{eq:X_r} holds,
then $X_r=\hat{X}_r$ for any $r\in\mc{R}$.
\end{lem}
\begin{proof}
Define $\mc{R}_1:=\{r\in\mc{R}: \hat{X}_r>X_{\rm r}\}$ and $\mc{R}_2:=\{r \in \mc{R}: \hat{X}_r\leq X_{\rm r}\}$.
Assume $\mc{R}_1\neq \emptyset$.
Letting $\mc{I}_0:= \{i \in \mc{I}: \hat{\lambda}^i>\lambda^i\}$, we have $\sum_{r\in\mc{R}_1}\hat{x}^i_r=D^i-\hat{x}^i_0-\sum_{r\in\mc{R}_2}\hat{x}^i_r\leq D^i-\hat{x}^i_0-\sum_{r\in\mc{R}_2}x^i_r$ for any $i \in \mc{I}_0$ from the first claim of Lemma~\ref{lem:ineq}.
Since $\hat{x}^i_0\geq 0,$ $D^i-\hat{x}^i_0-\sum_{r\in\mc{R}_2}x^i_r \leq D^i-\sum_{r\in\mc{R}_2}x^i_r$ for any $i \in \mc{I}_0$.
From the second claim of Lemma~\ref{lem:ineq}, $x^i_0=0$ for $i \in \mc{I}_0$ and $D^i-\sum_{r\in\mc{R}_2}x^i_r=D^i-\sum_{r\in\mc{R}_2}x^i_r-x^i_0=\sum_{r\in\mc{R}_1}x^i_r.$
Hence
\begin{equation}\label{eq:sumN0}
 \textstyle{\sum_{r\in\mc{R}_1} \hat{x}^i_r \leq \sum_{r\in\mc{R}_1} x^i_r,\quad \forall i \in \mc{I}_0.}
\end{equation}
On the other hand, from the first claim of Lemma~\ref{lem:ineq},
\begin{equation}\label{eq:sumnN0}
 \textstyle{\sum_{r\in\mc{R}_1} \hat{x}^i_r \leq \sum_{r\in\mc{R}_1} x^i_r,\quad \forall i\notin\mc{I}_0.}
\end{equation}
From~\eqref{eq:sumN0} and~\eqref{eq:sumnN0}, we have
\[
\begin{array}{ll}
 \sum_{r\in\mc{R}_1}\hat{X}_r \hs = \sum_{r\in\mc{R}_1} \sum_{i \in \mc{I}} \hat{x}^i_r\\
  \hs = \sum_{r\in\mc{R}_1} \sum_{i \in \mc{I}_0} \hat{x}^i_r+\sum_{r\in\mc{R}_1} \sum_{i\notin\mc{I}_0} \hat{x}^i_r\\
  \hs \leq \sum_{r\in\mc{R}_1} \sum_{i \in \mc{I}_0} x^i_r+\sum_{r\in\mc{R}_1} \sum_{i\notin\mc{I}_0} x^i_r\\
 \hs = \sum_{r\in\mc{R}_1}X_r.
\end{array}
\]
However, from the definition of $\mc{R}_1$, $\sum_{r\in\mc{R}_1}\hat{X}_r > \sum_{r\in\mc{R}_1}X_r$.
This is a contradiction and hence $\mc{R}_1=\emptyset$.
Similarly, $\hat{X}_r>X_r$ does never happen.
Therefore, the claim holds.
\end{proof}

Now we can prove Theorem~\ref{thm:uniqueness}.

\begin{proof}[Theorem~\ref{thm:uniqueness}]
If there exists a NE such that $X_r\leq \chi_r$ for some $r\in\mc{R}$, then every NE incurs no cost and hence the NE is essentially unique from Lemma~\ref{lem:delay}.
In the following, we assume that~\eqref{eq:X_r} holds.
If $\hat{\lambda}^i=\lambda^i$ for $i \in \mc{I}$, then $\hat{x}^i_r = x^i_r$ for any $i \in \mc{I}$ and $r\in\mc{R}$ from the first claim of Lemma~\ref{lem:ineq} and Lemma~\ref{lem:Xr}.
Consider the case where $\hat{\lambda}^i<\lambda^i$ for some $i \in \mc{I}$.
From the first claim of Lemma~\ref{lem:ineq} and Lemma~\ref{lem:Xr}, $\hat{x}^i_r=x^i_r=0$ or $\hat{x}^i_r<x^i_r$ for any $r\in\mc{R}$.
Assume that there exists $r\in\mc{R}$ such that $\hat{x}^i_r<x^i_r$.
Then $D^i=\sum_{r\in\mc{R}}\hat{x}^i_r+\hat{x}^i_0$.
From the second claim of Lemma~\ref{lem:ineq}, $\hat{x}^i_0=0$ and thus $D^i=\sum_{r\in\mc{R}}\hat{x}^i_r<\sum_{r\in\mc{R}}x^i_r\leq D^i$.
This is in contradiction.
Thus, in any case, $\hat{x}^i_r = x^i_r$ for any $i \in \mc{I}$ and $r\in\mc{R}$.
Then we have $\hat{x}^i_0=D^i-\sum_{r\in\mc{R}}\hat{x}^i_r=D^i-\sum_{r\in\mc{R}}x^i_r=x^i_0$.
Therefore, $\hat{x}=x$.
\end{proof}

\section{EQUILIBRIUM COMPUTATION}
\label{sec:alg}

This section proposes an algorithm to compute the NE.
Note that, the game $\mc{G}$ is not a potential game unlike non-atomic routing games.
Hence the algorithm that allows players to iteratively play best responses, and is guaranteed to converge to a NE for potential games~\cite[Theorem 6.4.3]{shoham2008multiagent}, may not converge for our game.

Instead, we consider a gradient-based algorithm, referred to as simultaneous improving response dynamics (SIRD) proposed in~\cite{Jacquot2018Analysis}, given by Algorithm~1.
In SIRD, $P_{\mc{X}^i}$ denotes the projection onto $\mc{X}^i$ and $\nabla_i J^i$ denotes the gradient of $J^i$ with respect to $x^i$.
The constants $\gamma>0$ and $\epsilon>0$ determine the step size and the stopping criterion, respectively.
SIRD can be regarded as an extended version of the gradient projection method~\cite[Sec.~2.3]{Bertsekas1998Nonlinear}, which was originally proposed for an optimization problem with a single decision maker.
In SIRD, every player simultaneously carries out each iteration of the gradient projection method.

\begin{algorithm}[th]
\caption{Simultaneous Improving Response Dynamics (SIRD)}
\begin{algorithmic}[1]
\REQUIRE{$x[0],\gamma,\epsilon$}
\ENSURE{$x$}
\STATE $k \leftarrow 0$
\REPEAT
\STATE $k\leftarrow k+1$
\FOR{$i=1,\ldots,N$}
\STATE $x^i[k] \leftarrow P_{\mc{X}^i}(x^i[k-1]-\gamma\nabla_i J^i(x[k-1]))$
\ENDFOR
\UNTIL{$\|x[k]-x[k-1]\|<\epsilon$}
\STATE $x\leftarrow x[k]$
\end{algorithmic}
\end{algorithm}

Our aim is to specify a condition with which SIRD converges to a NE.
In~\cite{Jacquot2018Analysis}, the convergence is proven under a mild condition for the case where there is no route corresponding to $x^i_0$ and the cost function $c^i_r$ is strictly increasing for the whole interval.
They have identified a condition with which the gradients become strongly monotone~\cite[Definition 2.3.1]{Facchinei2003Finite}, i.e., there exists $a>0$ such that
\begin{equation}\label{eq:str_jono}
 (x'-x)^{\sf T}(F(x')-F(x)) \geq a\|x'-x\|^2,
\end{equation}
for any $(x,x')\in\mc{X}\times\mc{X}$ with $F(x):=[\nabla_i J^i(x)]_{i \in \mc{I}}$.
From the strong monotonicity each iteration in SIRD becomes a contraction, which leads to the convergence from the standard fixed point theorem.

However, in our model, the condition~\eqref{eq:str_jono} does never hold because the aggregated gradient $F(x)$ can be constant at two different points.
Indeed, for any $x\in\mc{X}$ satisfying $X_r\leq \chi_r$ for any $r\in\mc{R}$ , we have $c_r(X_r)=c'_r(X_r)=0$ and hence $\nabla_iJ^i(x)=[\alpha^i\ 0\ \cdots\ 0]^{\sf T}$.
By choosing such $x$ and $x'$, we have $F(x')-F(x)=0$ for $x\neq x'$.
Thus it is clear that there does not exist $a>0$ that satisfies~\eqref{eq:str_jono}.
Therefore, the iteration in SIRD is not a contraction and the standard fixed point theorem cannot be used in our case.

Instead of the logic above, we employ a variant of the fixed point theorem adapted to averaged maps.
A vector-valued map $F:\mc{X}\to\bR^N$ is said to be \emph{nonexpansive} if
\[
 \|F(y)-F(x)\|^2 \leq \|y-x\|^2,\quad \forall (x,y)\in\mc{X}\times\mc{X}.
\]
Also, $F$ is said to be \emph{averaged} if there exist $\beta\in(0,1)$ and a nonexpansive map $T:\mc{X}\to\bR^N$ such that $F=(1-\beta)I+\beta T$.
It is known that iteration of an averaged map converges to a fixed point of the map~\cite[Theorem 2.1]{Byrne2003Unified}.
Furthermore, we need the notion of co-coercive~\cite[Definition 2.3.9]{Facchinei2003Finite} as an alternative to strong monotonicity:
A map $F:\mc{X}\to\mc{X}$ is said to be co-coercive if there exists $a>0$ such that
\[
 (x'-x)^{\sf T}(F(x')-F(x)) \geq a\|F(x')-F(x)\|^2
\]
for any $(x,x')\in\mc{X}\times\mc{X}$.
In what follows we show that if $F$ is co-coercive then the iteration in SIRD is averaged,
and hence it converges to a NE.

We start with showing that $F$ is co-coercive under the condition provided in~\cite{Jacquot2018Analysis}.
The following lemma holds.
\begin{lem}\label{lem:cocoercive}
Let Assumptions~\ref{assum:tech} and~\ref{assum:identical_intervals} hold.
If there exists $a>0$ such that
\begin{equation}\label{eq:cond_conv}
 2c'_r(X_r)\left(
 1-
 \left(
 \dfrac{c''_r(X_r)}{2c'_r(X_r)}
 \right)^2
 \|x_r\|^2
 \right)\geq a,\quad \forall x\in\mc{X}
\end{equation}
for any $r\in\mc{R}$, then $F$ is co-coercive.
Moreover, then $P_{\mc{X}} \circ (I-\gamma F)$ is averaged for $\gamma<2a$.
\end{lem}
\begin{proof}
Let $G$ denote the Jacobian of $F$ given by $\nabla^{\sf T} F(x)$.
Note that if $G(x)+G(x)^{\sf T}$ is positive semidefinite for any $x\in\mc{X}$ then there exists $b>0$ such that
\[
 y^{\sf T}G(x)y \geq b\|G(x)y\|^2,\quad \forall x\in\mc{X},\quad\forall y\in\bR^{N(R+1)},
\]
which implies that $F$ is co-coercive~\cite[Prop. 2.9.25 (a)]{Facchinei2003Finite}.
Thus it suffices to show that $G(x)+G(x)^{\sf T}$ is positive semidefinite for the former claim.

We re-index $G(x)$ to be block-diagonal by $G(x):={\rm diag}(G_0(x_0),G_1(x_1),\ldots,G_R(x_R))$ with
$G_r(x_r):=[\partial^2 J^i(x_r)/\partial x^i_r \partial x^j_r]_{i,j \in \mc{I}}.$
Then we have
\[
 G_r(x_r)+G_r(x_r)^{\sf T}=\left[\dfrac{\partial^2 J^i}{\partial x^i_r \partial x^j_r}(x_r) + \dfrac{\partial^2 J^j}{\partial x^j_r \partial x^i_r}(x_r)\right]_{i,j \in \mc{I}}.
\]
For $r=0$, every component is zero.
For $r=1,\ldots,R$,
\[
 \dfrac{\partial^2 J^i}{\partial x^i_r \partial x^j_r}(x_r) = \left\{
 \begin{array}{ll}
 2c'_r(X_r)+x^i_rc''_r(X_r) & {\rm if}\ i=j,\\
 c'_r(X_r)+x^i_rc''_r(X_r) & {\rm otherwise}.
 \end{array}
 \right.
\]

Denote the quadratic form by $q(y):=y^{\sf T}(G_r(x_r)+G_r(x_r)^{\sf T})y$
for a fixed $x_r$.
By omitting notation $r,$ $X_r,$ and $y$ for simplicity, we have
\[
 q = \sum_{i \in \mc{I}}2(2c'+x^ic'')y^2_i + \sum_{i \in \mc{I}} \sum_{j\neq i} (2c'+(x^i+x^j)c'')y_iy_j.
\]
Let $\zeta:=2c'$ and $\eta_{ij}:=(x^i+x^j)c''$.

If $\zeta\neq 0$, we have
\[
 \begin{array}{l}
 q =\sum_{i \in \mc{I}}(2\zeta_i+\eta_{ii})y^2_i+\sum_{i \in \mc{I}} \sum_{j\neq i}(\zeta+\eta_{ij})y_iy_j\\
 = \zeta \sum_{i \in \mc{I}}y^2_i + \zeta\left(\sum_{i \in \mc{I}}
    \left(
     1+\dfrac{\eta_{ii}}{2\zeta}
    \right)y_i
   \right)^2\\
  \displaystyle{
   + \hspace{-2mm} \sum_{i \in \mc{I},j\in\mc{I}}\left(\eta_{ij}+\zeta-\zeta
  \left(1+\dfrac{\eta_{ii}}{2\zeta} \right)\left(1+\dfrac{\eta_{jj}}{2\zeta} \right)
  \right)y_iy_j
  }\\
  \displaystyle{
   = \zeta \sum_{i \in \mc{I}}y^2_i + \zeta
  \sum_{i \in \mc{I}}
  \left(
   \left(1+\dfrac{\eta_{ii}}{2\zeta}\right) y_i
  \right)^2
  }\\
  \displaystyle{
   \quad  - \sum_{i \in \mc{I},j\in\mc{I}} \dfrac{\eta_{ii}\eta_{jj}}{4\zeta}y_iy_j.
  }
 \end{array}
\]
Thus the quadratic form can be decomposed as $q = y^{\sf T}(A_1+A_2+A_3)y$
with $A_1:=\zeta I,$ $A_2:=\zeta\xi\xi^{\sf T},$ and $A_3:= \eta\eta^{\sf T}/(4\zeta)$
where $\xi:=[1+\eta_{ii}/(2\zeta)]_{i \in \mc{I}}$ and $\eta:=[\eta_{ii}]_{i \in \mc{I}}.$
Note that
\[
 \sigma(A_1)=\{\zeta\},\ \sigma(A_2)=\{\zeta\|\xi\|^2\},\ \sigma(A_3)=\{\|\eta\|^2/(4\zeta)\}.
\]
and $\sigma(A_1+A_2)=\{\zeta,\zeta(1+\|\xi\|^2)\}.$
From Weyl's inequality~\cite[Theorem 4.3.1]{Horn2012Matrix}, we have
\[
\begin{array}{ll}
 \min \sigma(A_1+A_2-A_3) & \geq \min \sigma(A_1+A_2)-\max\sigma(A_3)\\
  & = \zeta -\|\eta\|^2/(4\zeta).
\end{array}
\]
Thus if $\zeta>0$ and $\zeta\geq\|\eta\|^2/(4\zeta)$, the latter of which is equivalent to~\eqref{eq:cond_conv}, then $G_r(x_r)+G_r(x_r)^{\sf T}$ is positive semidefinite.

On the other hand, if $\zeta=0$, then $X_r\in[0,\chi]$ and hence $c''(X_r)=0$.
Then $G_r(x_r)+G_r(x_r)^{\sf T}=0$ and hence the matrix is positive semidefinite for any $r\in\{0\}\cup\mc{R}$ and $x\in\mc{X}$.

After all, if~\eqref{eq:cond_conv} holds, then $G_r(x_r)+G_r(x_r)^{\sf T}$ is positive semidefinite.
Because $G(x)$ is block diagonal, we have $\sigma(G(x)+G(x)^{\sf T})=\cup_{r\in\{0\}\cup\mc{R}} \sigma(G_r(x_r)+G_r(x_r)^{\sf T}).$
Therefore, $G(x)+G(x)^{\sf T}$ is also positive semidefinite.

We show the latter claim.
Note that $I-\gamma F=(1-\frac{\gamma}{2a}\gamma)I+\frac{\gamma}{2a}(I-2aF).$
Now we have
\[
\begin{array}{l}
 \|x'-x\|^2-\|(I-2aF)(x')-(I-2aF)(x) \|^2\\
 = 4a(x'-x)^{\sf T}(F(x') - F(x)) - 4a^2\|F(x')-F(x)\|^2 \geq 0 
\end{array}
\]
because $F$ is co-coercive with modulus $a$.
Thus $I-2aF$ is nonexpansive.
This implies that $I-\gamma F$ is averaged for $\gamma<2a$.
Finally, because the composition of a projection and an averaged map is also averaged~\cite[Proposition 2.1]{Byrne2003Unified}, the latter claim holds.
\end{proof}

Lemma~\ref{lem:cocoercive} indicates that a NE can be computed using SIRD.
\begin{theorem}
Let Assumptions~\ref{assum:tech} and~\ref{assum:identical_intervals} hold.
If \eqref{eq:cond_conv} holds, then the sequence generated by SIRD with $\gamma<2a$ converges to a NE.
\end{theorem}
\begin{proof}
Note that the set of NE of the game $\mc{G}$ is the set of the solutions of the KKT condition~\eqref{eq:KKT}.
Also, those sets are the fixed points of $P_{\mc{X}}\circ (I-\gamma F)$.
Since a NE exists, a fixed point of $P_{\mc{X}}\circ (I-\gamma F)$ exists as well.
Hence, from Lemma~\ref{lem:cocoercive} and Theorem 2.1 in~\cite{Byrne2003Unified}, the claim holds.
\end{proof}

\section{NUMERICAL EXAMPLE}
\label{sec:num}
\subsection{Simulation Parameters}
We present a numerical example to discuss properties of the game based on the theoretical developments.
Consider a two-player game with two routes, i.e., $N=2$ and $R=2$.
The demands are set by $D^1=100$ and $D^2=150$.
We consider the delivery duration in a linear form of $T_r(X_r)=\mu_r X_r+\nu_r$ with $\mu_r>0$ and $\nu_r>0$.
Let $(\mu_1,\mu_2)=(0.3,0.5)$ and $(\nu_1,\nu_2)=(5,6)$.
The delayed delivery cost function is set to a quadratic function given by $\bar{c}^i(t)=(t-\tau^i)^2$ for $t\geq \tau^i$.
Note that those functions satisfy Assumption~\ref{assum:tech}.
This choice leads to $c^i_r(X_r)=(\nu_r X_r + \mu_r - \tau^i)^2$ for $X_r\geq \chi^i_r:= (\tau^i-\mu_r)/\nu_r$.

\subsection{Effectiveness of SIRD}
\begin{figure}[t]
  \centering
  \includegraphics[width=0.98\linewidth]{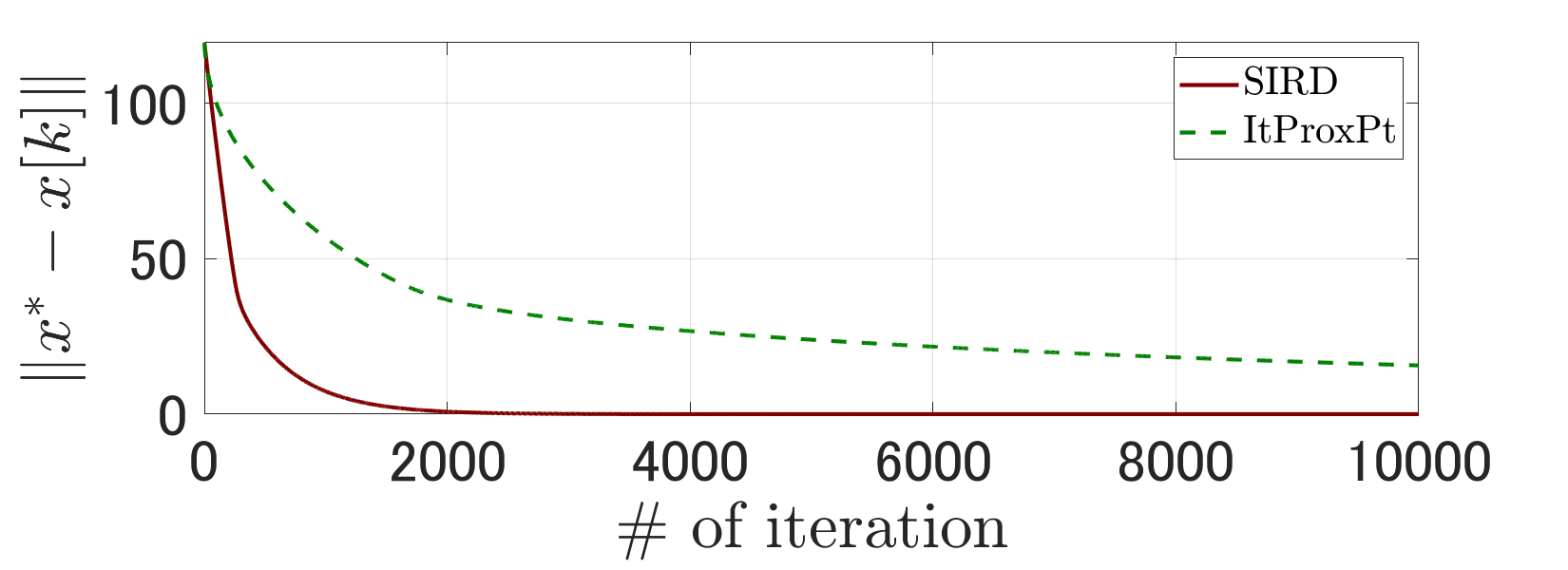}
  \caption{
  Convergence rate of SIRD given by Algorithm~1 with comparison to the distributed iterative proximal-point algorithm, referred to as ItProxPt, proposed in~\cite{Chen2014Autonomous}.
  }
  \label{fig:conv}
\end{figure}

We first observe the effectiveness of the proposed algorithm with comparison to the distributed iterative proximal-point algorithm, referred to as ItProxPt, proposed in~\cite{Chen2014Autonomous}.
The ItProxPt algorithm is a gradient-based algorithm as well, and Line 5 in Algorithm~1 is replaced with
$x^i[k] \leftarrow P_{\mc{X}^i}(x^i[k-1]-\gamma_{k-1}(\nabla _i J^i(x[k-1]) +\theta(x^i[k-1]-x^i[k-2])))$
where the time-varying parameter $\gamma_{k-1}$ is set to satisfy the Robbins-Monro condition~\cite[Chapter 2]{Borkar2009Stochastic}
and $\theta\in(0,1)$ is a constant.
Specifically, we set $\gamma_k=k^{-1/2}$ and $\theta=1/2$.
Let $(\tau^1,\tau^2)=(8,8)$, under which both players have the same delayed delivery cost function.
Fig.~\ref{fig:conv} draws the convergence rate of the algorithms for an instance with $(\alpha^1,\alpha^2)=(0.5,1.5)$, where the solid line corresponds to SIRD while the broken line corresponds to ItProxPt.
The horizontal and vertical lines describe the number of iteration and the norm of the error from the NE, respectively.
First, it is observed that SIRD successfully finds the NE.
Second, it is also observed that, SIRD converges to the NE very quickly compared with ItProxPt.
This arises from the property that the step size $\gamma$ can be fixed in SIRD while it is time-varying and converges to zero in ItProxPt.

\subsection{Inefficiency of NE}

\begin{figure}[t]
\centering
\subfloat[][]{\includegraphics[width=.93\linewidth]{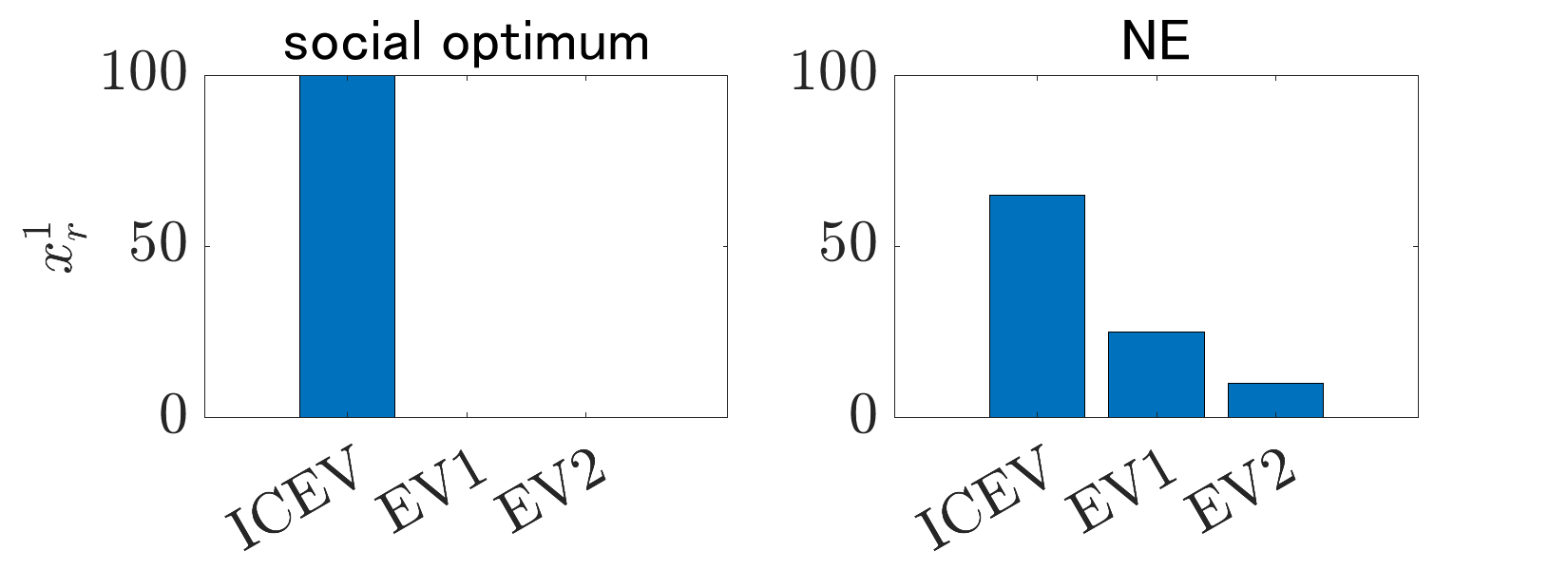}\label{subfig:low_pol}}\\
\subfloat[][]{\includegraphics[width=.93\linewidth]{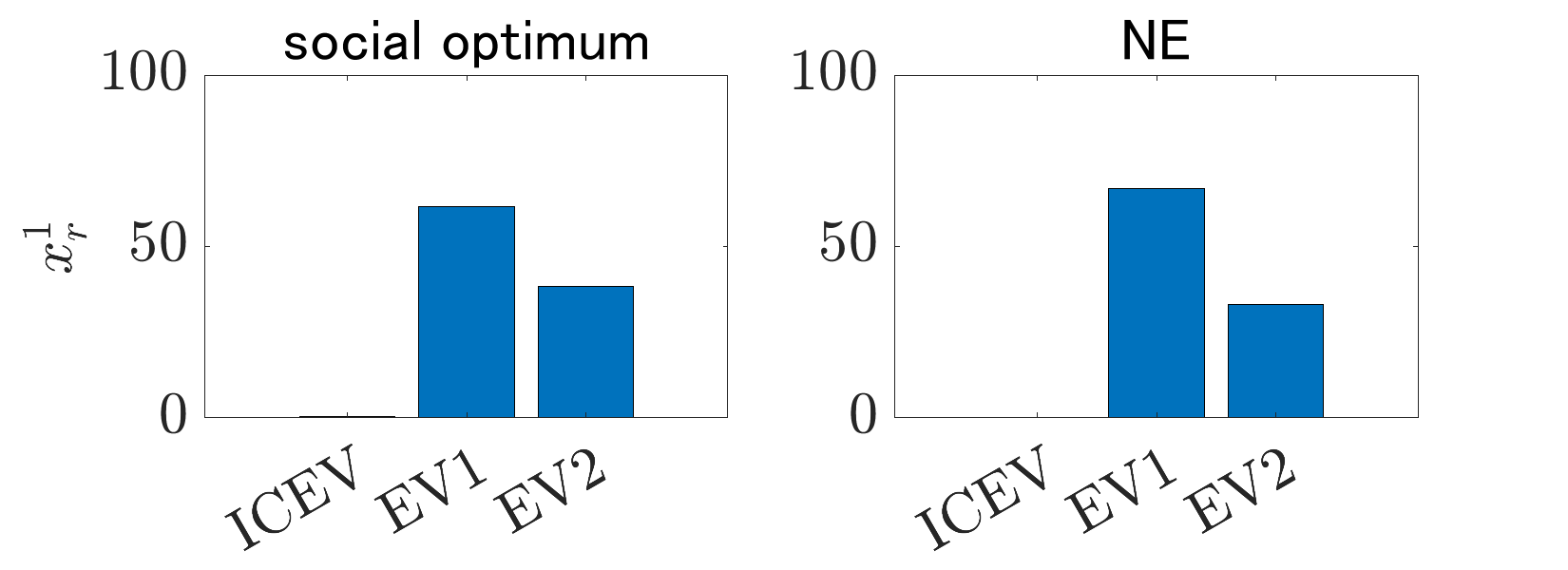}\label{subfig:high_pol}}
\caption[]{(a): Socially optimal strategy of player 1 and her equilibrium strategy with a low marginal pollution cost. (b): Those with a high marginal pollution cost.}
\end{figure}

\begin{figure}[t]
  \centering
  \includegraphics[width=0.98\linewidth]{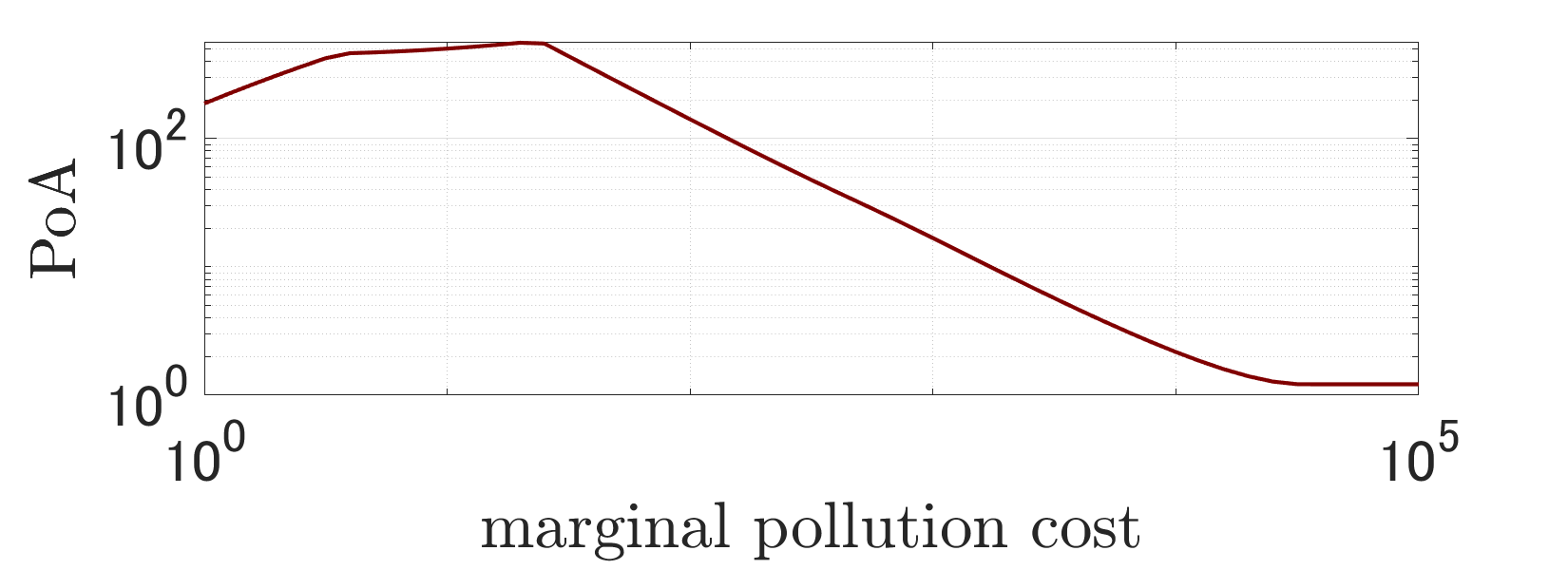}
  \caption{
   PoA with varying marginal pollution cost.
  }
  \label{fig:PoAcurve}
\end{figure}

We adopt the summation of the players' costs $\sum_{i\in\mc{I}}J^i(x)$ as the objective function.
Also, inefficiency of NE is measured by PoA, namely, the ratio between the worst social cost of a NE and that of an optimal outcome~\cite[Chap.~17]{Roughgarden2007Game}.
Fig.~\ref{subfig:low_pol} depicts the socially optimal profile and the unique NE of the instance considered in the previous subsection.
The PoA is 187.6.
As suggested by the extremely large PoA, the social optimum and the NE are considerably different.
It can be observed that player 1 fully uses ICEVs at the social optimum, while she assigns moderate shares to EVs at the NE.
A possible approach to reduce inefficiency is to increase the marginal pollution cost.
Let $(\alpha^1,\alpha^2)=(0.5C_{\alpha},1.5C_{\alpha})$ with a varying parameter $C_{\alpha}\geq1$.
Fig.~\ref{fig:PoAcurve} draws the PoA where the horizontal line is $C_{\alpha}$.
It is observed that the PoA when $C_{\alpha}=10^5$ is about 1.2.
The social optimum and the NE with high marginal pollution cost are depicted in Fig.~\ref{subfig:high_pol}.

\subsection{Multiple NE Case}
\begin{figure}[t]
  \centering
  \includegraphics[width=0.95\linewidth]{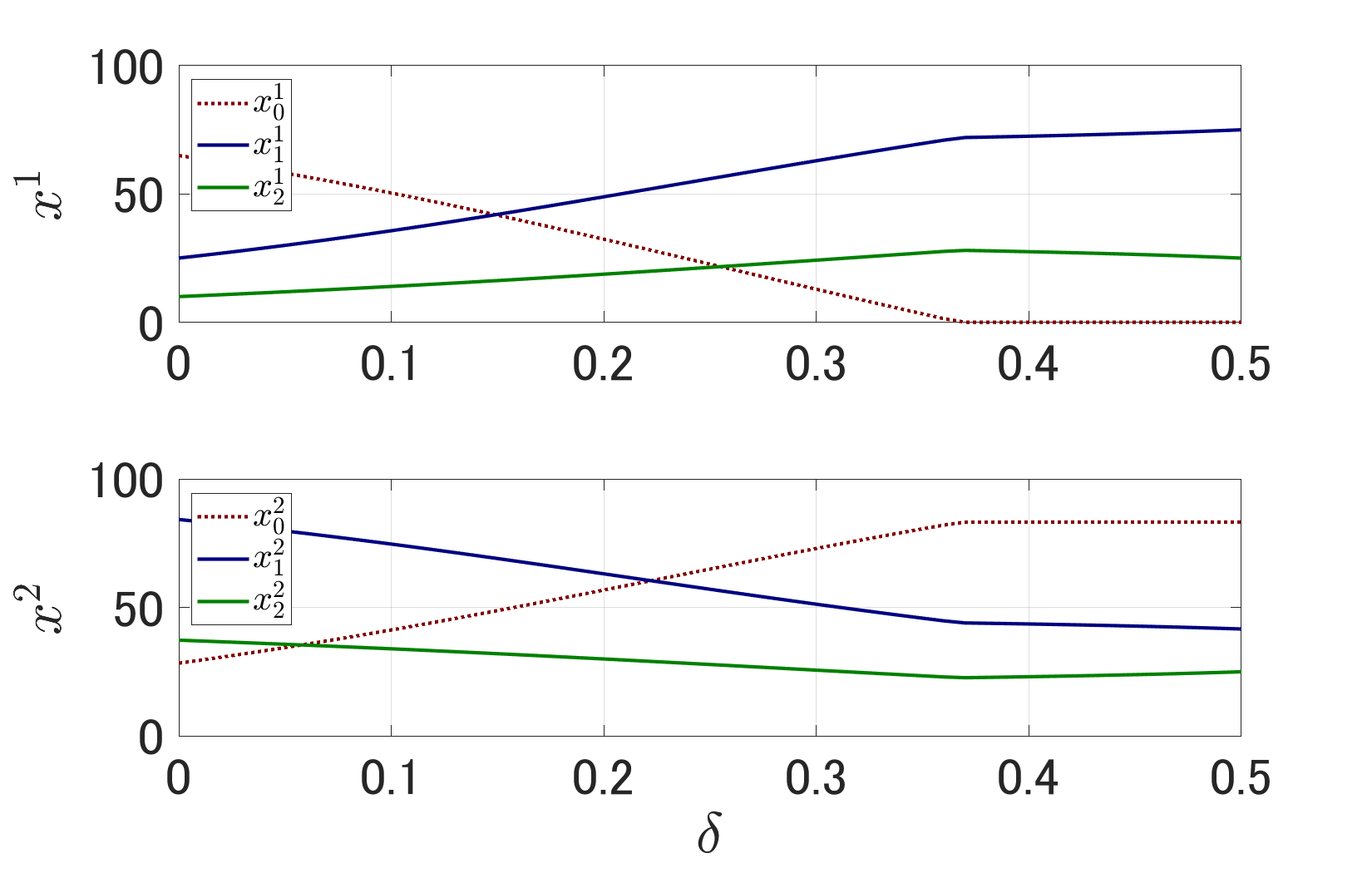}
  \caption{
   NE vs $\delta$, which prolongs the required delivery time of player 1.
  }
  \label{fig:NEvsdelta}
\end{figure}

Next, we investigate a condition for existence of multiple NE.
Introduce a parameter $\delta\geq0$ that prolongs the required delivery time of player 1 by $\tau^1=8+\delta$.
In fact, as $\delta$ increases, the gap between the cost functions of the players increases, which leads to existence of multiple NE as expected in Sec.~\ref{subsec:ex}.
It can be confirmed that the NE is unique when $\delta \in [0,0.5]$.
Fig.~\ref{fig:NEvsdelta} depicts the equilibrium flow profile with varying $\delta$.
When $\delta$ is large, player 1 uses only EVs due to less strict requirement.
When $\delta>0.5$ the gap between the cost functions are such huge that multiple NE arises.

\begin{figure}[t]
  \centering
  \includegraphics[width=0.95\linewidth]{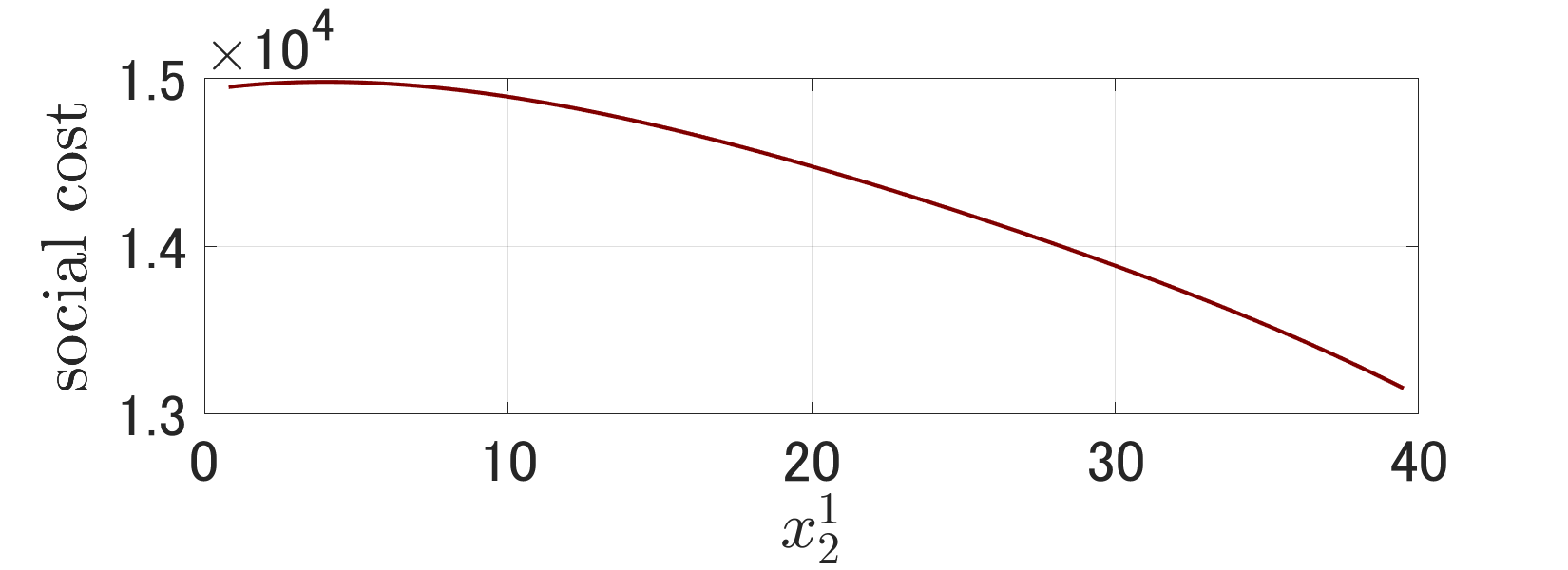}
  \caption{
  Social cost at the NE vs $x^1_2$ in the case with multiple NE.
  }
  \label{fig:SocCosvsx12}
\end{figure}

Finally, we observe inefficiency of multiple NE.
Let $\delta=0.7$ and then there exist multiple NE.
It can be confirmed that $x^1=(0,D^1-x^1_2,x^1_2)$ can be a NE with the best response of player 2 for $x^1_2\in[0.8,39.5]$.
Fig.~\ref{fig:SocCosvsx12} depicts the social cost corresponding to each NE.
It is observed that the social cost is minimized when $x^1_2$ is maximized.
Note that $\mu_1<\mu_2$ and $\nu_1<\nu_2$, and hence the route 2 is the most time-consuming route.
This result suggests that it is socially desirable that players whose time requirement is less strict than the others put all demand to the most time-consuming routes.

\section{CONCLUSION}
\label{sec:conc}

The paper has formulated a green routing game, where the cost is composed of pollution and delayed delivery with interaction caused by congestible nature at public charging stations.
In contrast to standard routing games, the game admits multiple NE.
It has been shown that the NE is unique when the delayed delivery cost functions are identical with respect to players.
Moreover, a distributed algorithm has been proposed.
Numerical results have revealed several properties of the green routing game.
In particular, it is suggested that high marginal pollution cost can reduce inefficiency of selfish routing.
The obtained results provide insights for mechanism design of internalizing congestion and environment externality.
Future work includes theoretical analysis of PoA.


\bibliographystyle{IEEEtran}
\bibliography{sshrrefs}

\begin{thebibliography}{10}
\providecommand{\url}[1]{#1}
\csname url@samestyle\endcsname
\providecommand{\newblock}{\relax}
\providecommand{\bibinfo}[2]{#2}
\providecommand{\BIBentrySTDinterwordspacing}{\spaceskip=0pt\relax}
\providecommand{\BIBentryALTinterwordstretchfactor}{4}
\providecommand{\BIBentryALTinterwordspacing}{\spaceskip=\fontdimen2\font plus
\BIBentryALTinterwordstretchfactor\fontdimen3\font minus
  \fontdimen4\font\relax}
\providecommand{\BIBforeignlanguage}[2]{{%
\expandafter\ifx\csname l@#1\endcsname\relax
\typeout{** WARNING: IEEEtran.bst: No hyphenation pattern has been}%
\typeout{** loaded for the language `#1'. Using the pattern for}%
\typeout{** the default language instead.}%
\else
\language=\csname l@#1\endcsname
\fi
#2}}
\providecommand{\BIBdecl}{\relax}
\BIBdecl

\bibitem{European2021National}
{European Environment Agency}, ``National emissions reported to the {UNFCCC}
  and to the {EU} greenhouse gas monitoring mechanism,'' 2021, [Online].
  Available:
  \url{https://www.eea.europa.eu/data-and-maps/data/national-emissions-reported-to-the-unfccc-and-to-the-eu-greenhouse-gas-monitoring-mechanism-17}.

\bibitem{European2020Stepping}
{European Commission}, ``Stepping up {E}urope’s 2030 climate ambition,''
  2020, [Online]. Available:
  \url{https://eur-lex.europa.eu/legal-content/EN/TXT/?uri=CELEX:52020DC0562}.

\bibitem{Yong2015Review}
J.~Y. Yong, V.~K. Ramachandaramurthy, K.~M. Tan, and N.~Mithulananthan, ``A
  review on the state-of-the-art technologies of electric vehicle, its impacts
  and prospects,'' \emph{Renewable and Sustainable Energy Reviews}, vol.~49,
  pp. 365--385, 2015.

\bibitem{Bektacs2011Pollution}
T.~Bekta{\c{s}} and G.~Laporte, ``The pollution-routing problem,''
  \emph{Transportation Research Part B: Methodological}, vol.~45, no.~8, pp.
  1232--1250, 2011.

\bibitem{Schneider2014Electric}
M.~Schneider, A.~Stenger, and D.~Goeke, ``The electric vehicle-routing problem
  with time windows and recharging stations,'' \emph{Transportation Science},
  vol.~48, no.~4, pp. 500--520, 2014.

\bibitem{Goeke2015Routing}
D.~Goeke and M.~Schneider, ``Routing a mixed fleet of electric and conventional
  vehicles,'' \emph{European Journal of Operational Research}, vol. 245, no.~1,
  pp. 81--99, 2015.

\bibitem{Kocc2016Green}
{\c{C}}.~Ko{\c{c}} and I.~Karaoglan, ``The green vehicle routing problem: {A}
  heuristic based exact solution approach,'' \emph{Applied Soft Computing},
  vol.~39, pp. 154--164, 2016.

\bibitem{Shen2019Optimization}
Z.-J.~M. Shen, B.~Feng, C.~Mao, and L.~Ran, ``Optimization models for electric
  vehicle service operations: {A} literature review,'' \emph{Transportation
  Research Part B: Methodological}, vol. 128, pp. 462--477, 2019.

\bibitem{Asghari2021Green}
M.~Asghari and S.~{Al-e-hashem}, ``Green vehicle routing problem: {A}
  state-of-the-art review,'' \emph{International Journal of Production
  Economics}, vol. 231, 2021.

\bibitem{Roughgarden2007Game}
N.~Nisan, T.~Roughgarden, \'{E}va Tardos, and V.~Vazirani, Eds.,
  \emph{Algorithmic Game Theory}.\hskip 1em plus 0.5em minus 0.4em\relax
  Cambridge University Press, 2007.

\bibitem{Krichene2014Stackelberg}
W.~Krichene, J.~D. Reilly, S.~Amin, and A.~M. Bayen, ``Stackelberg routing on
  parallel networks with horizontal queues,'' \emph{IEEE Trans. Autom.
  Control}, vol.~59, no.~3, pp. 714--727, 2014.

\bibitem{Lazar2021Routing}
D.~A. Lazar, S.~Coogan, and R.~Pedarsani, ``Routing for traffic networks with
  mixed autonomy,'' \emph{IEEE Trans. Autom. Control}, vol.~66, no.~6, pp.
  2664--2676, 2021.

\bibitem{Wu2021Value}
M.~Wu, S.~Amin, and A.~E. Ozdaglar, ``Value of information in {Bayesian}
  routing games,'' \emph{Operations Research}, vol.~69, no.~1, pp. 148--163,
  2021.

\bibitem{Orda1993Competitive}
A.~Orda, R.~Rom, and N.~Shimkin, ``Competitive routing in multiuser
  communication networks,'' \emph{IEEE/ACM Transactions on Networking}, vol.~1,
  no.~5, pp. 510--521, 1993.

\bibitem{Chen2014Autonomous}
H.~Chen, Y.~Li, R.~H. Louie, and B.~Vucetic, ``Autonomous demand side
  management based on energy consumption scheduling and instantaneous load
  billing: {A}n aggregative game approach,'' \emph{IEEE trans. Smart Grid},
  vol.~5, no.~4, pp. 1744--1754, 2014.

\bibitem{Jacquot2018Analysis}
P.~Jacquot, O.~Beaude, S.~Gaubert, and N.~Oudjane, ``Analysis and
  implementation of an hourly billing mechanism for demand response
  management,'' \emph{IEEE Trans. Smart Grid}, vol.~10, no.~4, pp. 4265--4278,
  2018.

\bibitem{Jacquot2018Routing}
P.~Jacquot and C.~Wan, ``Routing game on parallel networks: {T}he convergence
  of atomic to nonatomic,'' in \emph{IEEE Conference on Decision and Control
  (CDC)}, 2018, pp. 6951--6956.

\bibitem{Byrne2003Unified}
C.~Byrne, ``A unified treatment of some iterative algorithms in signal
  processing and image reconstruction,'' \emph{Inverse Problems}, vol.~20, pp.
  103--120, 2003.

\bibitem{Al2021Charging}
B.~Al-Hanahi, I.~Ahmad, D.~Habibi, and M.~A. Masoum, ``Charging infrastructure
  for commercial electric vehicles: {C}hallenges and future works,'' \emph{IEEE
  Access}, vol.~9, pp. 121\,476--121\,492, 2021.

\bibitem{Work2010Traffic}
D.~B. Work, S.~Blandin, O.-P. Tossavainen, B.~Piccoli, and A.~M. Bayen, ``A
  traffic model for velocity data assimilation,'' \emph{Applied Mathematics
  Research eXpress}, vol. 2010, no.~1, pp. 1--35, 2010.

\bibitem{Polson2017Deep}
N.~G. Polson and V.~O. Sokolov, ``Deep learning for short-term traffic flow
  prediction,'' \emph{Transportation Research Part C: Emerging Technologies},
  vol.~79, pp. 1--17, 2017.

\bibitem{Tomaszewska2019Lithium}
A.~Tomaszewska, Z.~Chu, X.~Feng, S.~O'Kane, X.~Liu, J.~Chen, C.~Ji, E.~Endler,
  R.~Li, L.~Liu \emph{et~al.}, ``Lithium-ion battery fast charging: {A}
  review,'' \emph{ETransportation}, vol.~1, 2019.

\bibitem{Laraki2019Mathematical}
R.~Laraki, J.~Renault, and S.~Sorin, \emph{Mathematical Foundations of Game
  Theory}, ser. Universitext.\hskip 1em plus 0.5em minus 0.4em\relax Springer,
  2019.

\bibitem{BoyVan}
S.~Boyd and L.~Vandenberghe, \emph{Convex Optimization}.\hskip 1em plus 0.5em
  minus 0.4em\relax Cambridge University Press, 2004.

\bibitem{shoham2008multiagent}
Y.~Shoham and K.~Leyton-Brown, \emph{Multiagent systems: Algorithmic,
  game-theoretic, and logical foundations}.\hskip 1em plus 0.5em minus
  0.4em\relax Cambridge University Press, 2008.

\bibitem{Bertsekas1998Nonlinear}
D.~P. Bertsekas, \emph{Nonlinear Programming}.\hskip 1em plus 0.5em minus
  0.4em\relax Athena Scientific, 1998.

\bibitem{Facchinei2003Finite}
F.~Facchinei and J.-S. Pang, \emph{Finite-Dimensional Variational Inequalities
  and Complementarity Problems}.\hskip 1em plus 0.5em minus 0.4em\relax
  Springer, 2003.

\bibitem{Horn2012Matrix}
R.~A. Horn and C.~R. Johnson, \emph{Matrix Analysis}, 2nd~ed.\hskip 1em plus
  0.5em minus 0.4em\relax Cambridge University Press, 2012.

\bibitem{Borkar2009Stochastic}
V.~S. Borkar, \emph{Stochastic Approximation: A Dynamical Systems
  Viewpoint}.\hskip 1em plus 0.5em minus 0.4em\relax Springer, 2009.

\end{thebibliography}

\end{document}